\documentclass[aps,pra,
reprint,
superscriptaddress,
showpacs,
nofootinbib,
 amsmath,amssymb,
 aps,
]{revtex4-1}

\usepackage{graphicx}% Include figure files
\usepackage{epstopdf}
\usepackage{enumerate}
\usepackage{dcolumn}% Align table columns on decimal point
\usepackage{bm}% bold math
\usepackage{color}

\newcommand{\vect}[1]{\mbox{\boldmath $ #1 $}}

\begin{document}

%\preprint{APS/123-QED}

\title{Counter-diabatic vortex pump in spinor Bose--Einstein condensates}% Force line breaks with \\
%\thanks{A footnote to the article title}%

\author{T. Ollikainen}
%\footnote{Present adddress, QCD Labs, Department of Applied Physics, Aalto University, Aalto 00076, Finland}
\affiliation{QCD Labs, COMP Centre of Excellence, Department of Applied Physics, Aalto University, P.O. Box 13500, FI-00076 Aalto, Finland}
\author{S. Masuda}
\affiliation{QCD Labs, COMP Centre of Excellence, Department of Applied Physics, Aalto University, P.O. Box 13500, FI-00076 Aalto, Finland}
\author{M. M\"ott\"onen}
\affiliation{QCD Labs, COMP Centre of Excellence, Department of Applied Physics, Aalto University, P.O. Box 13500, FI-00076 Aalto, Finland}
\affiliation{University of Jyv\"askyl\"a, Department of Mathematical Information Technology, P.O. Box 35, FI-40014 University of Jyv\"askyl\"a, Finland}
\author{M. Nakahara}
%\affiliation{Department of Physics, Shanghai University, 200444 Shanghai, People’s Republic of China}
\affiliation{QCD Labs, COMP Centre of Excellence, Department of Applied Physics, Aalto University, P.O. Box 13500, FI-00076 Aalto, Finland}
\affiliation{Research Center for Quantum Computing and Department of Physics, Kinki University, Higashi-Osaka, 577-8502, Japan}

% \email{nakahara@math.kindai.ac.jp}
\date{\today}% It is always \today, today,
             %  but any date may be explicitly specified

\begin{abstract}
Topological phase imprinting is a well-established technique for deterministic vortex creation in spinor Bose--Einstein condensates of alkali metal atoms. It was recently shown that counter-diabatic quantum control may accelerate vortex creation in comparison to the standard adiabatic protocol and suppress the atom loss due to nonadiabatic transitions. Here we apply this technique, assisted by an optical plug, for vortex pumping to theoretically show that sequential phase imprinting up to 20 cycles generates a vortex with a very large winding number. Our method significantly increases the fidelity of the pump for rapid pumping compared to the case without the counter-diabatic control, leading to the highest angular momentum per particle reported to date for the vortex pump. Our studies are based on numerical integration of the three-dimensional multi-component Gross--Pitaevskii equation which conveniently yields the density profiles, phase profiles, angular momentum, and other physically important quantities of the spin-1 system. Our results motivate the experimental realization of the vortex pump and studies of the rich physics it involves. % Dependency of the pumping efficiency on the initial density profile of the condensate is also studied.

%\begin{description}
%\item[Usage]
%Secondary publications and information retrieval purposes.
%\item[PACS numbers]
%May be entered using the \verb+\pacs{#1}+ command.
%\item[Structure]
%You may use the \texttt{description} environment to structure your abstract;
%use the optional argument of the \verb+\item+ command to give the category of each item.
%\end{description}
\end{abstract}

% PACS, the Physics and Astronomy
                             % Classification Scheme.
%\keywords{Suggested keywords}%Use showkeys class option if keyword
                              %display desired
\maketitle

%\tableofcontents

\section{Introduction}\label{sec:1}
When the Bose--Einstein condensate (BEC) of alkali metal atoms was discovered in 1995~\cite{bec1,bec2}, some of the immediate issues to be clarified were whether this system exhibits quantum coherence and whether it shows superfluidity, similar to that of superfluid $^4$He. Quantized vortices are manifestations of these properties and their realization has been one of the main research topics of experimental BEC physics since its discovery~\cite{Fetter:2009}. In general, the existence of stable vortices and other topological objects is attributed to an order parameter manifold with nontrivial homotopy groups supporting these objects~\cite{topo_mermin,gtp}.

Methods to create vortices in BECs by so-called topological phase imprinting have been proposed in Refs.~\cite{top1,top2,top3,top4,toprev,Kawaguchi1,Kawaguchi2} and these proposals were later experimentally realized by several groups~\cite{exp1,exp15,exp16,exp2,exp25,exp3,exp4}. In addition to vortex creation, an extension of this method into a non-trivially three-dimensional scenario~\cite{Pietila2009} has been used in the experimental creation of a Dirac monopole in the synthetic magnetic field of a spinor BEC~\cite{Ray2014}. Other methods to create vortices include the utilization of moving laser beams~\cite{Matthews1999,Madison2000,Inouye2001}, rotating trap potentials~\cite{Hodby2001}, Laguerre--Gauss beams~\cite{Andersen2006}, and merging multiple condensates~\cite{Scherer2007}. The advantage of the topological phase imprinting method is that it creates vortices deterministically at a desired location and almost all atoms in the condensate acquire the desired angular momentum.

The operating principle of a vortex pump~\cite{Pietila2007} is based on sequential application of the topological phase imprinting protocol, thus increasing the winding number by $2F$ for each vortex pumping cycle, where $F$ is the quantum number of the hyperfine spin. Conventionally, various vortex pumping methods resorts to adiabatic control of the system to imprint local Berry phase~\cite{Xu2008,Xu2010,Kuopanportti2013}. Hence, rapid pumping gives rise to errors owing to unwanted nonadiabatic transitions, eventually leading to the degradation of the pump. On the other hand, a vortex with a large winding number is dynamically unstable into splitting into multiple single-quantum vortices~\cite{Kawaguchi:2004}, motivating faster vortex pumping. The stability of the large-winding-number states has been studied~\cite{Kuopanportti2010}, as well as their splitting dynamics~\cite{Kuopanportti2010_2}.

Importantly, the counter-diabatic (CD) protocol~\cite{cd1,cd2}, which is sometimes referred to as assisted adiabatic population transfer or shortcut to adiabaticity~\cite{review,review2}, can generate the same state as the corresponding adiabatic dynamics in shorter time. Therefore, it can be utilized to overcome several problems in adiabatic quantum control, for example, in cases in which the population transfer efficiency is limited by decoherence, three-body losses, and external noise in the control parameters of, e.g., isolated atoms and molecules~\cite{cd1,cd3,Masuda2014}, spin chain systems \cite{Campo2012,Takahashi2013}, Bose--Einstein condensates~\cite{Bason}, and electron spin of a single nitrogen-vacancy center in a diamond~\cite{Zhang2013}. Very recently, the CD protocol was also found to speed up the topological phase imprinting method~\cite{ourcd}. However, vortex pumping using the CD protocol has not been reported to date.

The purpose of this paper is to demonstrate that the CD quantum control, assisted by an optical plug, can accelerate the vortex pump, and hence create a large-winding-number state in a short time. This also serves to reduce the atom loss from the trap. We consider a spin-1 BEC consisting of $^{87}$Rb atoms in an optical trap with a three-dimensional quadrupole field present. The parameters in the simulations are set to experimentally feasible values according to Refs.~\cite{Ray2014,Ray2015}. In addition, we introduce an optical plug along the symmetry axis of the condensate to prevent transitions between the hyperfine spin states during the fast magnetic field ramp. We study the fidelity of the vortex creation and the state of the condensate for up to 20 vortex pumping cycles by numerically integrating the three-dimensional multi-component Gross--Pitaevskii equation.% To the best of our knowledge, this is the first application of the counter-adiabatic protocol to vortex pumping.

% We plot the atom number and density profile of each hyperfine state and also plot the phase profile of the condensate wave function. Vortex pumping was first proposed in \cite{Pietila2007}, which required a quadrupole field and a hexapole field. The axes of these fields must be alinged exactly in the same place for vortex stability, which is extremely difficult in practice. Our proposal makes use of a time-dependent transformation so that the required magnetic field is realized by a single quadrupole field, which makes physical implementation considerably easier.

This paper is organized as follows. In Sec.~\ref{sec:2}, we analyze single vortex creation ramps with the CD quantum control. We consider four cases: linear and nonlinear ramps with and without the CD field. We compare the performance of these schemes to show that the nonlinear ramp with the CD protocol assisted by an optical plug yields the highest fidelity. Furthermore, the effect of the condensate aspect ratio on the vortex creation fidelity is studied. In Sec.~\ref{sec:3}, we apply the results obtained in Sec.~\ref{sec:2} to vortex pumping. Detailed density and phase profiles are studied for nonlinear ramps with the CD field. Section~\ref{sec:4} is devoted to conclusions.

\section{Topological vortex imprinting with counter-diabatic field}\label{sec:2}
\subsection{Mean-field theory and topological vortex imprinting}
The mean-field order parameter of the spin-1 BEC is represented in the basis of the $z$-quantized spin states, $\left\{ \left| +1 \right>,\left| 0 \right>,\left| -1 \right>\right\}$, as $\Psi({\vect r},t)=\left( \psi_{+1}({\vect r},t),\psi_{0}({\vect r},t),\psi_{-1}({\vect r},t)\right)^T_z$, where the subscript in each spinor component denotes the magnetic quantum number along $z$. Furthermore, we write $\psi_k({\vect r},t)=\sqrt{n_k({\vect r},t)}\exp\left[i\phi_k({\bm  r},t)\right]$, where $n_k({\vect r},t)$ is the particle density and $\phi_k({\vect r},t)$ is the phase of the spinor component $k$. The dynamics of the mean-field order parameter are solved in three dimensions employing the Gross--Pitaevskii (GP) equation
\begin{align}
i\hbar \partial_t\Psi({\vect r},t) = &\left[-\frac{\hbar^2}{2m}\nabla^2 + V({\vect r}) + c_0\Psi({\vect r},t)^\dagger\Psi({\vect r},t) \right.\nonumber \\
&+ c_2\Psi({\vect r},t)^\dagger{\vect F}\Psi({\vect r},t)\cdot{\vect F}\nonumber\\
&\left.\vphantom{\frac{\hbar^2}{2m}}+g_F\mu_{\text{B}}{\vect B}({\vect r},t) \cdot {\vect F}\right]\Psi({\vect r},t),\label{eq:hamiltonian}
\end{align}
where the external optical potential is given by $V({\vect r})=V_{\text{opt}}(\rho,z) + V_{\text{plug}}(\rho)$, the harmonic part is $V_{\text{opt}}(\rho,z)=m\left(\omega_\rho^2\rho^2+\omega_z^2z^2\right)/2$, and $V_{\text{plug}}(\rho)=A_{\text{plug}}\exp(-\rho^2/\rho_{\text{plug}}^2)$ is the optical plug potential defined in the cylindrical coordinate system $(\rho,\varphi,z)$. Furthermore, ${\vect B}({\vect r},t)$ is the external time-dependent magnetic field, and ${\vect F}=(F_x,F_y,F_z)$ is a vector composed of the standard dimensionless spin-1 matrices. The constants $c_0=4\pi\hbar^2(a_0+2a_2)/(3m)$ and $c_2=4\pi\hbar^2(a_2-a_0)/(3m)$ are the coupling constants related to the density--density and spin--spin interactions~\cite{Ho:1998}, respectively, $g_F$ is the Land\'e $g$ factor, and $\mu_{\text{B}}$ is the Bohr magneton. For $F=1$ $^{87}$Rb atoms, the $s$-wave scattering lengths are given by $a_0=5.387$~nm and $a_2=5.313$~nm~\cite{vanKempen:2002}, the atomic mass by $m=1.443\times10^{-25}$~kg, and $g_F=-1/2$. The number of atoms is set to $N~=~2.1\times10^5$ throughout the simulations and the optical plug parameters are set to $\rho_{\text{plug}}= 1.80~\mu$m and $A_{\text{plug}}=6.6\times10^{-30}$~J in the cases in which the optical plug is employed.

In the topological vortex imprinting scheme considered, we employ an external magnetic field consisting of a three-dimensional quadrupole field with an additional bias field along $z$ defined as
\begin{align}
{\vect B}({\vect r},t) &= b_{\text{q}}(x\hat{\vect x} + y\hat{\vect y} - 2z\hat{\vect z}) + B_0(t)\hat{\vect z}\nonumber\\
%&=b_{\text{q}}(x\hat{\vect x} + y\hat{\vect y}) + B_z(z,t)\hat{\vect z}\nonumber\\
&=b_{\text{q}}(\rho\cos\varphi\hat{\vect x} + \rho\sin\varphi\hat{\vect y}) + B_z(z,t)\hat{\vect z},\label{eq:bfield}
\end{align}
where $b_{\text{q}}$ is the strength of the gradient field and $B_0(t)$ is the bias field component along $z$. We have further defined $B_z(z,t) =B_0(t) -2b_{\text{q}}z$. 

Let us consider an atom at a point ${\vect r}$ at time $t$. In the presence of the magnetic field $\vect{B}$, considering only the Zeeman term in the Hamiltonian, $\mathcal{H}_{\text{Z}}=g_F \mu_{\text{B}} {\vect B}\cdot {\vect F}$, we have three eigenstates corresponding to the weak-field-seeking state (WFSS) which has the highest energy, the neutral state (NS) with zero energy, and the strong-field-seeking state (SFSS) which has the lowest energy. In the $z$-quantized basis, these eigenstates are represented as
\begin{align}
\left|{\text{WFSS}}\right> & \hat{=} \frac{1}{2B}\begin{pmatrix}B-B_z\\-\sqrt{2}b_{\text{q}}\rho e^{i\varphi}\\\left(B+B_z\right)e^{2i\varphi}\end{pmatrix}_z,\label{eq:wfss}\\
\left|{\text{NS}}\right> & \hat{=} \frac{1}{\sqrt{2}B}\begin{pmatrix}-b_{\text{q}}\rho \\\sqrt{2}B_ze^{i\varphi}\\b_{\text{q}}\rho e^{2i\varphi}\end{pmatrix}_z,\label{eq:ns}\\
\left|{\text{SFSS}}\right> & \hat{=} \frac{1}{2B}\begin{pmatrix}B+B_z\\\sqrt{2}b_{\text{q}}\rho e^{i\varphi}\\\left(B-B_z\right)e^{2i\varphi}\end{pmatrix}_z,\label{eq:sfss}
\end{align} %The sign in the complex exponential is different compared to the case with the two-dimensional quadrupole field due to the difference in the winding number of the quadrupole field at $z=0$ evaluated along a circle centered at the origin~\cite{ourcd}. 
where $B(\rho,z,t)=\sqrt{b_{\text{q}}^2\rho^2+B_z(z,t)^2}$ is the total magnetic field strength. In contrast to the earlier work in Ref.~\cite{ourcd}, here we consider an optically trapped condensate and the magnetic field is primarly used only to control the spin state. Hence, we are free to choose also the magnetically untrapped SFSS as the initial state. Under a strong bias field we have $B\approx B_z$ at the location of the atoms, and consequently the state is approximately $(1,0,0)^{T}_z$. As the bias field is adiabatically ramped to a large negative value, such that $B_z\approx-B$, the state transforms into approximately $(0,0,e^{2i\varphi})^{T}_z$. The appearance of the azimuthal dependence in the phase factor can be attributed to the accumulation of the Berry phase during the adiabatic change of the bias field~\cite{Berry1984}.

While the bias field is inverted, there appears a space-time point where the magnetic field and hence Zeeman energy gap between the eigenstates vanishes. Near this point, transitions from SFSS to NS and WFSS take place. However, atoms in the SFSS are naturally repelled from the location of the magnetic field zero, which suppresses such transitions. Due to this reason, we take SFSS as the initial state in most of our simulations. Additionally, we employ the optical plug in some of the simulations to further diminish the atom loss and to stabilize the resulting multi-quantum vortex~\cite{Kuopanportti2010}.

\subsection{Counter-diabatic field for a three-dimensional quadrupole field}%Employing the magnetic field of Eq.~(\ref{eq:bfield}) in the vortex creation scheme requires adiabatic control of the bias field. Rapid inversion of the bias field gives rise to nonadiabatic transitions, hence lowering the fidelity of the vortex creation.
In an adiabatic process, the populations in the instantaneous eigenstates of the time-dependent Hamiltonian remain constant. If the system is not driven slow enough, transitions between the quantum states takes place. Counter-diabatic protocol can be used to designate an auxiliary term in the Hamiltonian to overcome the requirement of slow driving. In the CD protocol, the auxiliary term removes the nonadiabatic transitions and hence generates the same final state as the adiabatic process.~\cite{cd1,review2}

The dynamics of a spin-1 system in the presence of a changing magnetic field are described by the Zeeman Hamiltonian $\mathcal{H}_{\text{Z}}$. The auxiliary Hamiltonian provided by the CD scheme in this case is given by $\mathcal{H}_{\text{CD}} = g_F\mu_{\text{B}}{\vect B}_{\text{CD}}\cdot \vect F$, where the so-called CD field reads ${\vect B}_{\text{CD}}=\hbar{\vect B}\times\partial_t{\vect B}/(g_F\mu_{\text{B}}\left|{\vect B}\right|^2)$.~\cite{cd2}

In the following, we may approximate $B_z(z,t)\approx B_z(0,t)=B_0(t)$. Thus the CD field in the topological vortex imprinting method, according to protocol in Eq.~(\ref{eq:bfield}), assumes the form
\begin{equation}
{\vect B}_{\text{CD}}({\vect r},t) = \frac{\hbar \dot{B}_0(t)b_{\text{q}}}{g_F \mu_{\text{B}} B^2(\rho,0,t)}(y\hat{\vect x}-x\hat{\vect y}),\label{eq:cd}
\end{equation}
where $\dot{B}_0(t)$ denotes the temporal derivative of the bias field. Hence, the required external magnetic field including the CD field is thus ${\vect B} + {\vect B}_{\text{CD}}$.  It is also possible to solve the CD field without setting $z=0$, but it turns out to be of complex form and not likely conveniently realizable within the current experimental equipment. Furthermore, we make the approximation $\rho= \rho_0$ in the denominator of Eq.~(\ref{eq:cd}) due to the requirement $\nabla\cdot({\vect B}+{\vect B}_{\text{CD}})=0$ imposed by the Maxwell's equations.

Implementing the modified magnetic field ${\vect B} + {\vect B}_{\text{CD}}$ is experimentally challenging, since it would require two separate sets of exactly aligned quadrupole coils. In order to make this scheme experimentally convenient, we consider a time-dependent unitary transformation $U(t)=\exp\left[ -i\alpha(t) F_z \right]$, similar to that in Ref.~\cite{ourcd}, which introduces a rotation of an angle 
\begin{equation}
\alpha(t) = \arctan\left(\frac{\left|{\vect B}_{\text{CD}}({\vect r},t) \right|}{b_{\text{q}}\rho} \right)
\end{equation}
about the $z$-axis. The order parameter transforms into $\Psi'({\vect r},t)=U(t)\Psi({\vect r},t)$. In the beginning and at the end the vortex creation process of duration $T$, we have $|{\vect B}|\gg|{\vect B}_{\text{CD}}|$, and consequently $\alpha(0)=\alpha(T)\approx0$, i.e., $U(0)=U(T)\approx I$, where $I$ is the identity. Hence, the order parameters $\Psi'({\vect r},t)$ and $\Psi({\vect r},t)$ coincide in the beginning and at the end of the ramp, and we may steer the system with the effective Hamiltonian for $\Psi'({\vect r},t)$ to achieve the desired CD dynamics.

\begin{figure}[t]
\centering
\includegraphics[width=0.45\textwidth]{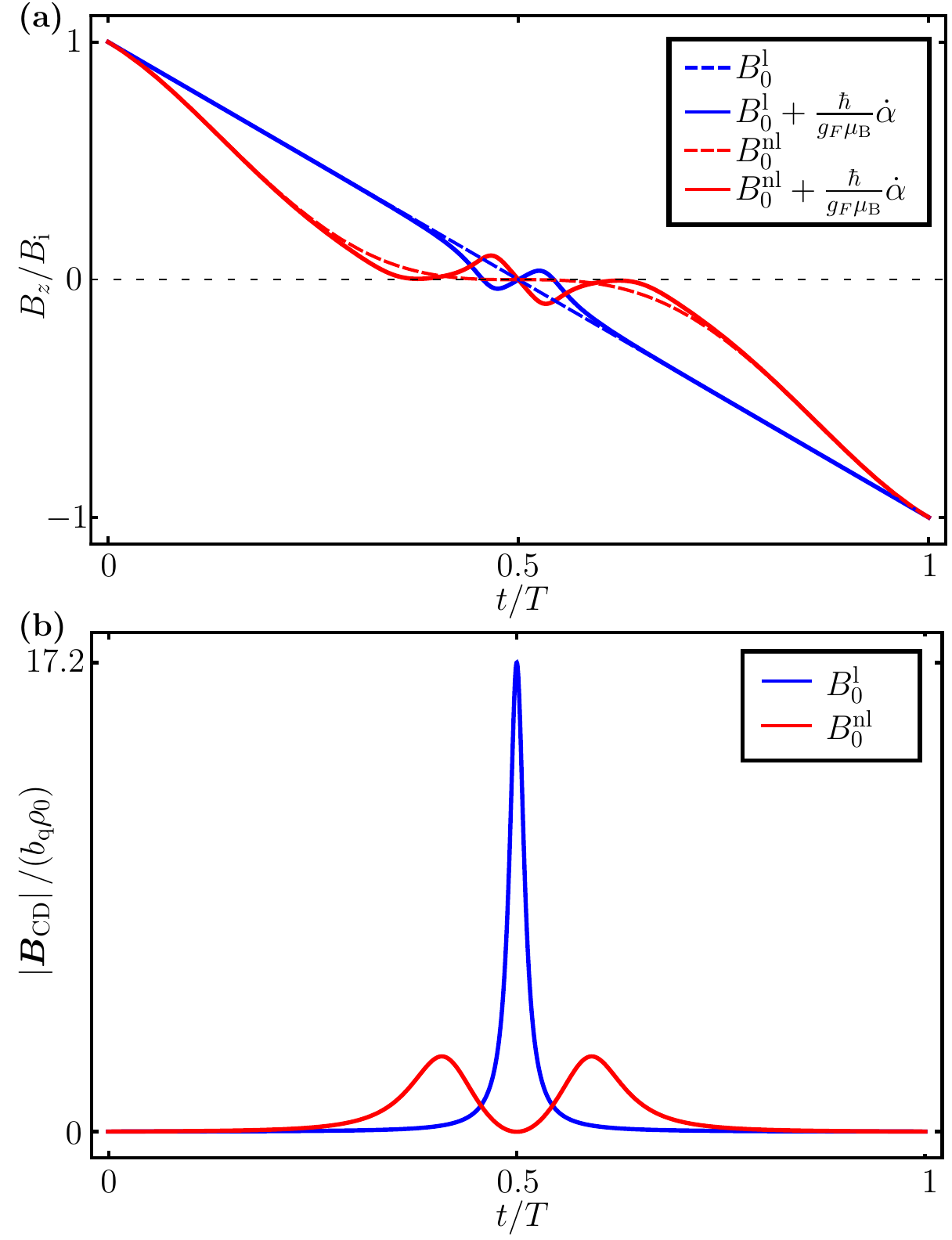}
\caption{(Color online) (a) Bias field component along $z$ and (b) the strength of the CD field as functions of time $t$. We choose $B_{\text{i}}=0.5$~G, $\rho_0=3~\mu\text{m}$, $b_{\text{q}}=3.675$~G/cm, and the total ramp time $T=1.16~\text{ms}$. The blue and red lines correspond to the linear and nonlinear ramps, respectively.}
\label{fig:rampfunctions}
\end{figure}

The Zeeman part of the Hamiltonian~(\ref{eq:hamiltonian}) transforms for $\Psi'({\vect r},t)$ into $g_F\mu_\text{B}U(t)({\vect B}+{\vect B}_{\text{CD}})\cdot{\vect F}U(t)^\dagger$. This gives rise to a rotation of the magnetic field by an angle $\alpha(t)$ about $z$ axis. Furthermore, the transformed Hamiltonian includes an additional term $-i\hbar U(t)\partial_t U(t)^\dagger=\hbar\dot{\alpha}(t)F_z$, which we take into consideration by adding a magnetic field $\hbar\dot{\alpha}(t)/(g_F\mu_{\text{B}})$ along $z$. The resulting magnetic field assumes the form
\begin{align}
\tilde{\vect B}({\vect r},t) =& b_\text{q}\rho\sqrt{1+\left[\frac{\hbar\dot{B}_0(t)}{g_F\mu_\text{B}B^2(\rho_0,0,t)}\right]^2}\left( \hat{\vect x} + \hat{\vect y}- 2 \hat{\vect z} \right)\nonumber\\ 
& + \left[B_0(t) +\frac{\hbar}{g_F\mu_{\text{B}}} \dot{\alpha}(t)\right]\hat{\vect z}.
\label{eq:cdbfieldrotated}
\end{align}
As is evident from the above equation, also the gradient field along $z$ is affected by the CD field in our simulations. As a result, only one set of quadrupole coils is required to implement the resulting magnetic field.

In the vortex creation scheme employed here, the bias field $B_z$ is ramped from a large positive value to a large negative value while keeping the quadrupole field strength $b_\text{q}$ constant. We consider two different ramping functions
\begin{equation}
B_0^{\text{l}}(t) = (1-2t/T)B_{\text{i}},
\end{equation}
and
\begin{equation}
B_0^{\text{nl}}(t) = g(t)(1-2t/T)B_{\text{i}},
\end{equation}
where $T$ is the duration of the ramp, $t\in\left[0,T\right]$, and $B_{\text{i}}$ is the initial strength of the magnetic field. Ramping functions $B_0^{\text{l}}$ and  $B_0^{\text{nl}}$ are henceforth referred to as \emph{linear} and \emph{nonlinear} ramping functions, respectively. For the nonlinear ramp, we set 
\begin{equation}
g(t) = \frac{1}{2}\left\{1-\cos\left[\frac{2\pi\left(t-T/2\right)}{T} \right]\right\}. 
\end{equation}

The ramp functions $B_0^{\text{l}}$ and $B_0^{\text{nl}}$, their transformed counterparts, and the strength of the CD field for both ramp functions are shown in Fig.~\ref{fig:rampfunctions}. The nonlinear ramp requires a weaker CD field strength compared with the linear ramp, hence reducing the amount of electric current needed in the quadrupole coils.

\subsection{Fidelity of vortex creation using counter-diabatic quantum control}

\begin{figure}[h]
\centering
\includegraphics[width=0.38\textwidth]{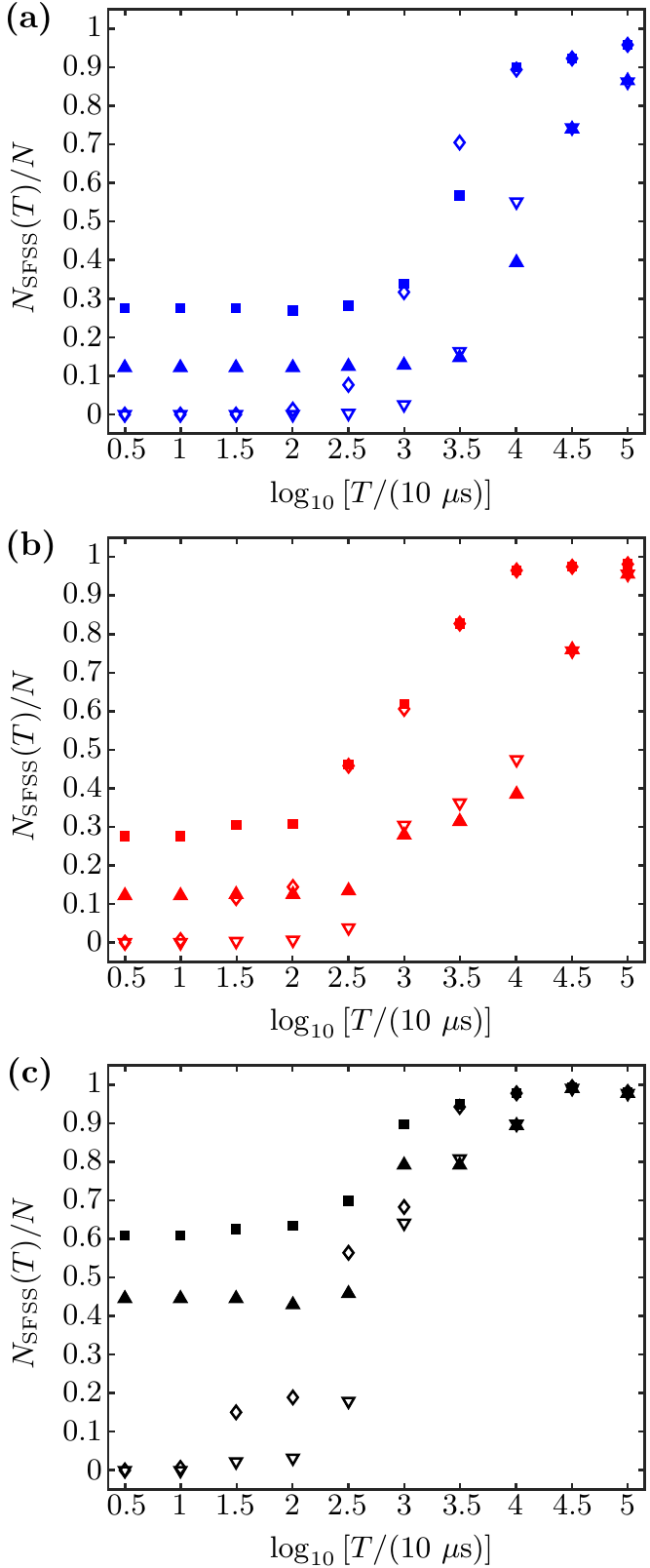}
\caption{(Color online) Fidelity of the vortex creation as a function the ramp time. We study (a) linear ramps, (b) nonlinear ramps, and (c) nonlinear ramps with the optical plug. Filled squares and empty diamonds correspond to the case with and without the CD field, respectively, for $(\omega_\rho,\omega_z)=2\pi\times(24.8,124)$~Hz, and filled upward-pointing and empty downward-pointing triangles correspond to the case with and without the CD field, respectively, for $(\omega_\rho,\omega_z)=2\pi\times(124,164)$~Hz.}
\label{fig:timedependence}
\end{figure}

We define the fidelity of vortex creation as the fraction of atoms trapped in the SFSS, $N_{\text{SFSS}}(t)$, with respect to the conserved total atom number, $N$, after a single creation ramp. Initially all atoms reside in the SFSS, i.e., $N_{\text{SFSS}}(0)=N$. In an ideal case, without any non-adiabatic excitations, we would have $N_{\text{SFSS}}(T)=N$. The initial and final values for the bias field are set to $B_0(0)=0.5$~G and $B_0(T)=-0.5$~G, respectively. The quadrupole field strength is linearly ramped on in the beginning of the simulations and off at the end of the simulations with the bias field kept constant. The ramp times for setting the quadrupole field on and off are both roughly 0.06 ms. Hence, we can evaluate $N_{\text{SFSS}}=\int\!d{\vect r}\, |\psi_{+1}|^2$ at the beginning and $N_{\text{SFSS}}=\int\!d{\vect r}\, |\psi_{-1}|^2$ at the end of the simulations. We consider two cases for the optical trapping potentials: $(\omega_\rho,\omega_z)=2\pi\times(124,164)$~Hz and $(\omega_\rho,\omega_z)=2\pi\times(24.8,124)$~Hz. These correspond to slighly oblate and moderately oblate three-dimensional condensates, respectively. We choose the parameter $\rho_0$ to approximately correspond to the radial coordinate of the density maximum of the condensate at the beginning of the simulations: for slightly (moderately) oblate condensates we set $3.0$ $(6.7)$~$\mu$m in the case without the plug and $6.0$ $(13.4)$~$\mu$m with the plug. The optimal value for $\rho_0$ varies depending on the ramp time $T$~\cite{ourcd}, and it is not fully optimized for all ramp times.

The resulting fidelity of the vortex creation is presented in Fig.~\ref{fig:timedependence} as a function of the ramp time $T$. The effect of the CD field is most dramatic for brief ramps, i.e., when the bias field is controlled in a nonadiabatic manner. For $T= 10$~$\mu$s, both linear and nonlinear ramps without the optical plug give essentially the same result: the CD field improves the fidelity from almost zero to 0.1 and 0.3 for slightly and moderately oblate condensates, respectively. At longer times, $T > 10~\mu\text{s}$, nonlinear ramp yields slightly higher fidelity since the zero of the field moves more slowly in the condensate region. 

The optical plug further enhances the fidelity for all ramp times, since the atoms are repelled from the path of the field zero. Indeed, the fidelity is very close to unity at $T>100~\text{ms}$ if the optical plug is employed, regardless whether the CD field is used or not. Here, also the effect of the CD field is negligibly small because $\dot{\alpha}(t)$ is small.

\subsection{Effect of condensate aspect ratio on the vortex creation fidelity}

\begin{figure}[]
\centering
\includegraphics[width=0.39\textwidth]{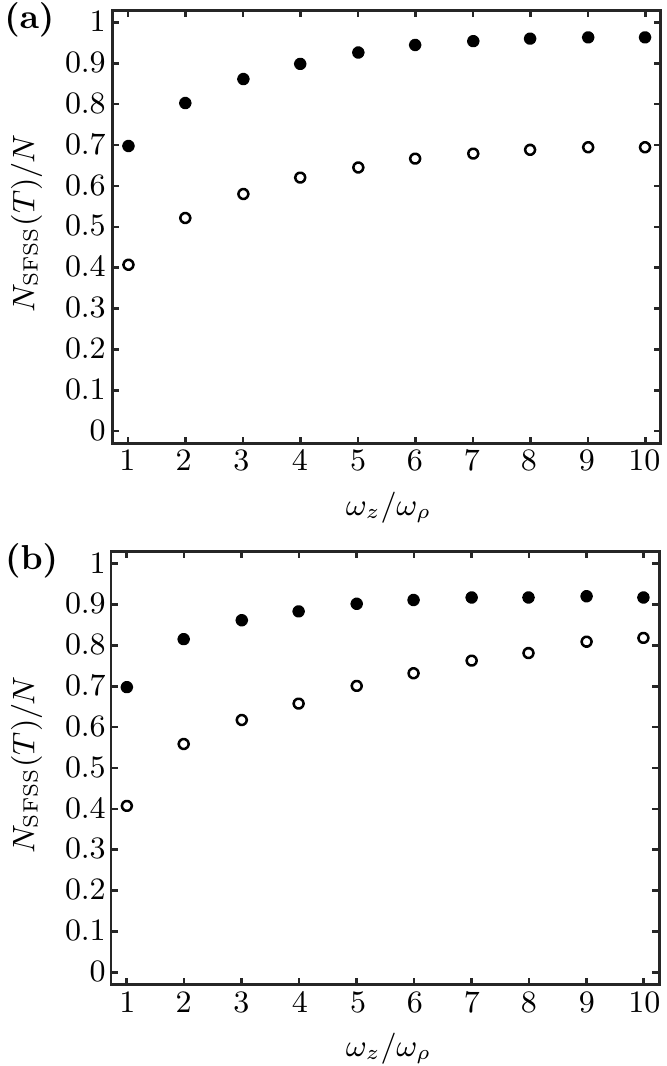}
\caption{Fidelity of the vortex creation as a function of the aspect ratio of the trap frequencies $\omega_z/\omega_\rho$ in the case of nonlinear ramps and optical plug, for (a) $\omega_\rho=2\pi\times124$~Hz and for (b) $\omega_z=2\pi\times124$~Hz. Filled and empty circles correspond to the cases with and without the CD field, respectively. The ramp time $T=5.78$~ms and the CD parameter $\rho_0=\sqrt{(2\pi\times124~\text{Hz})/\omega_\rho}\times 6$~$\mu$m.}
\label{fig:thicknessdependence}
\end{figure}

In the derivation of the CD field of Eq.~(\ref{eq:cd}) we have set $z=0$. Hence, the CD field is most beneficial for highly oblate rather than spherical or prolate BECs. In Fig.~\ref{fig:thicknessdependence}, we study the effect of the aspect ratio of the BEC cloud on the vortex creation fidelity by varying the aspect ratio of the trap frequencies $\omega_z/\omega_\rho$.

We find that, in general, the vortex creation process is more precise with more oblate condensates. This is the case regardless whether we choose to include the CD field in the creation process or not. Without the CD field however, the fidelity saturates well below unity if $\omega_\rho$ is kept constant. In contrast, if $\omega_\rho$ is varied, the effective condensate width changes. Due to the increase in the effective width, also the region in which the vortex creation is nearly adiabatic is increased, providing higher vortex creation fidelity.

\pagebreak

\onecolumngrid
\begin{center}
\begin{figure}[h]
\centering
\includegraphics[width=\textwidth]{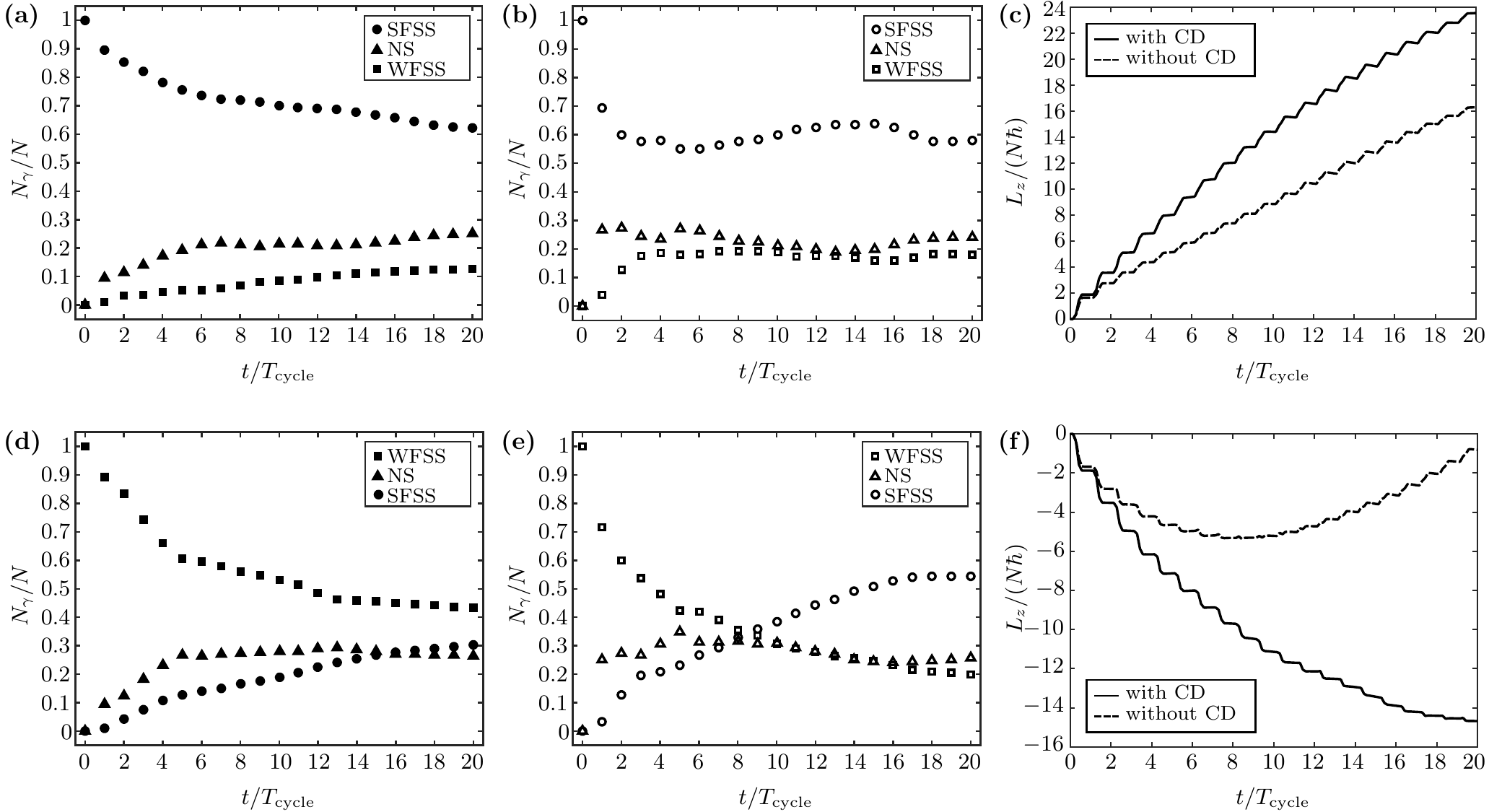}
\caption{(a,b,d,e) Particle number $N_\gamma$ for the Zeeman eigenstate $\gamma\in\left\{\text{WFSS},\text{NS},\text{SFSS} \right\}$ as a function of time (a,d) with and (b,e) without the CD field. (c,f) Orbital angular momentum along $z$ as a function of time. The initial state is (a--c) SFSS and (d--f) WFSS.}
\label{fig:pumpingboth}
\end{figure}
\end{center}
\twocolumngrid

\section{Vortex pumping with counter-diabatic quantum control}\label{sec:3}

Let us apply the CD control to vortex pumping, in which we cyclically increase the angular momentum of the vortex hosted in the BEC. Here, we choose $\omega_z/\omega_\rho=~5$ with $\omega_z=2\pi\times124$~Hz, $\rho_0=13.4~\mu\text{m}$, and employ the optical plug to ensure that, with the CD field, roughly 90\% of the atoms will remain in the initial state after the first vortex pumping cycle. Furthermore, the optical plug serves to prevent the created multi-quantum vortex from splitting into multiple single-quantum vortices~\cite{Kuopanportti2010_2}. Here, we consider vortex pumping up to 20 cycles. Each cycle consists of the following sequential steps: %The amplitude of the optical plug is set to $A_{\text{plug}}=6.6\times10^{-30}$~J and the radius $r_{\text{plug}}\approx1.8$~$\mu$m.

\begin{enumerate}[I.]
\item The quadrupole field strength $b_{\text{q}}$ is linearly set from zero to 3.675~G/cm in $T_1= 0.06$~ms while keeping the bias field $B_0$ fixed at $B_{\text{i}}=0.1$~G.\label{it:1}
\item The bias field $B_0$ is ramped to $-B_\text{i}$ using the nonlinear ramp with the CD field in $T_2=1.16$~ms.\label{it:2}
\item The quadrupole field strength $b_q$ is linearly ramped from 3.675~G/cm to zero in $T_3=0.06$~ms.\label{it:3}
\item The bias field $B_0$ is linearly rotated to its initial state in $T_4=0.51$~ms while keeping its magnitude constant. Here we utilize an additional bias field in the $y$ direction.\label{it:4}
\end{enumerate}

Steps \ref{it:1}, \ref{it:3}, and \ref{it:4} have essentially no effect in the relative populations of the Zeeman eigenstates. Here we have chosen $B_{\text{i}}=0.1$~G, instead of $B_{\text{i}}=0.5$~G, to reduce the ramp time by a factor of five compared with Fig.~\ref{fig:timedependence}, while the amount of nonadiabatic excitations remain constant\footnote{We may use $B_{\text{i}}=0.1$~G because, for the chosen ramp time, $\alpha(t)$ is negligibly small, and consequently $\Psi'({\vect r},t)\approx\Psi({\vect r},t)$ in the beginning and at the end of Step~\ref{it:2}.}. We numerically verified that the simulations give equivalent results if the additional bias field is applied along $x$ instead of $y$ in Step~\ref{it:4}. The total duration of one cycle in this scheme is only $T_{\text{cycle}}= 1.79$ ms. This is significantly faster than vortex pumping relying on standard adiabatic dynamics~\cite{Pietila2007,Xu2008,Xu2010,Kuopanportti2010,Kuopanportti2013}. 

\begin{figure}[t]
\centering
\includegraphics[width=0.45\textwidth]{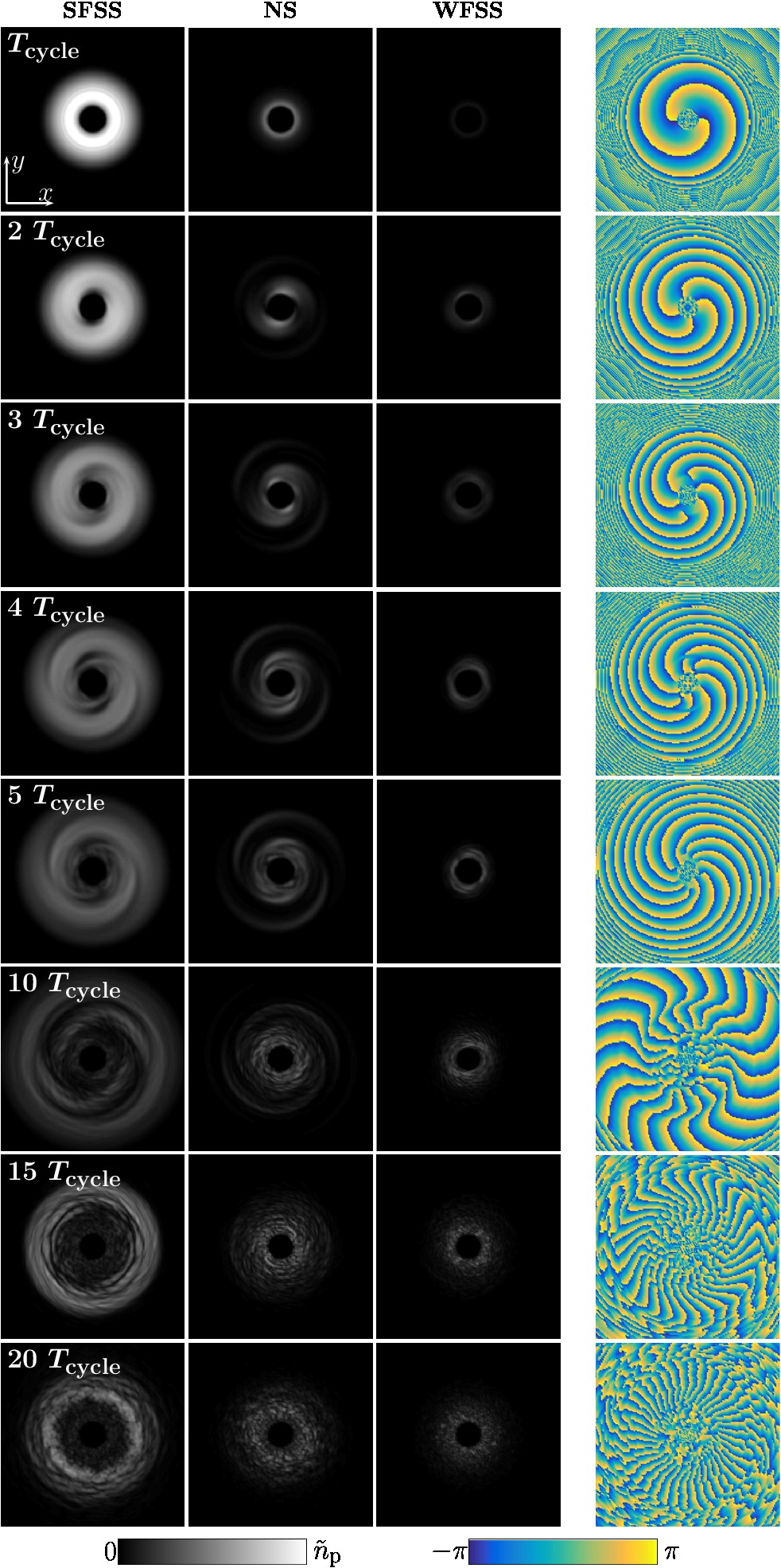}
\caption{(Color online) Particle densities of the SFSS, NS, and WFSS integrated along the $z$-axis, corresponding to Fig.~\ref{fig:pumpingboth}(f). The field of view in each panel is $30\times30$~$\mu\text{m}^2$ and the maximum particle density is $\tilde{n}_{\text{p}}=1.52\times10^{11}$~cm$^{-2}$. The rightmost column shows the phase of SFSS in the $z=0$ plane.}
\label{fig:wavefunction1}
\end{figure}

The particle numbers for the Zeeman eigenstates and the $z$ components of the condensate orbital angular momentum, $L_z=-i\hbar\int\!{\rm d}{\vect r}\,\Psi^\dagger\left(x\partial_y - y\partial_x \right) \Psi$, at different stages of the vortex pumping process are shown in Fig.~\ref{fig:pumpingboth}. In the beginning of the simulation and at the end of Step~\ref{it:4}, the particle numbers for each cycle can be conveniently evaluated as $N_\text{WFSS}=\int\!{\rm d}{\vect r}\,|\psi_{-1}|^2$, $N_\text{NS}=\int\!{\rm d}{\vect r}\,|\psi_{0}|^2$, and $N_\text{SFSS}=\int\!{\rm d}{\vect r}\,|\psi_{+1}|^2$. After 20 cycles, the CD protocol yields an orbital angular momentum of approximately $-14N\hbar$ and $23N\hbar$ with the initial conditions corresponding to WFSS and SFSS, respectively. This is a clear improvement compared with the case without the CD protocol, for which approximately $-6N\hbar$ is reached at the ninth cycle and $16N\hbar$ is reached at the end of the twentieth cycle, for the initial conditions corresponding to WFSS and SFSS, respectively. In an ideal case, a 20-cycle vortex pumping process yields orbital angular momentum of $\pm40N\hbar$, since each cycle provides an additional $\pm2N\hbar$. This value is not achieved due to the nonadiabatic transitions occurring at every cycle between the different Zeeman eigenstates. Thus, the evolving state is a combination of WFSS, NS, and SFSS. 

For WFSS as an initial state, in the case without the CD field, the orbital angular momentum starts increasing roughly at the ninth cycle when SFSS becomes the dominant state. For SFSS as an initial state, it remains as the dominant state throughout the 20-cycle process. The strong-field-seeking state yields higher fidelity in vortex pumping because in comparison to WFSS, the condensate accumulates further away from the field zero into the region where the spin rotations caused by the external magnetic field are more adiabatic.

The order parameters at various instants of time in the course of vortex pumping are shown in Fig.~\ref{fig:wavefunction1}. Here, the initial state is SFSS and the CD protocol is applied. The phase information reveals the precise accumulation of the winding number during the pumping process although the angular momentum does not exactly increase by $2N\hbar$ per cycle. The spiraling phase pattern is attributed to the additional breathing of the condensate during the vortex creation process. 

\section{Conclusions}\label{sec:4}
We have numerically studied topological vortex imprinting and vortex pumping in spinor BECs of $^{87}$Rb atoms aided by counter-diabatic quantum control. The employed CD control can be achieved with a single set of quadrupole field coils and the simulation parameters are chosen to be experimentally feasible. We demonstrate that the counter-diabatic field can be used to reduce the atom loss in the topological vortex imprinting process also in the case of an optical plug. The highest fidelity in the nonadiabatic regime in the vortex creation process is achieved in our simulations with nonlinear ramps employing both the optical plug and the CD scheme. We also find that the more oblate the condensate, the higher the vortex creation fidelity. Importantly, we show that the CD control and the optical plug can be used to accelerate the vortex pumping process in comparison to the standard adiabatic protocol. This speed-up leads to the highest angular momentum per particle reported to date for the vortex pump.% We find that if the initial state is chosen as SFSS rather than WFSS, the absolute value for the orbital angular momentum is higher in the resulting state. 

The experimental realization of the vortex pump remains a milestone to be achieved in the studies of topological defects in spinor BECs. Our results show that the strict requirement of adiabaticity in conventional vortex pumping can be relaxed by employing the CD scheme. Faster pumping may also provide stabilization against the splitting of vortices with large winding number during the pumping process.

\begin{acknowledgments}
We thank David Hall for insightful discussions. This  research  has  been  supported  by  the  Academy  of Finland  through  its  Centres  of  Excellence  Program  (grant nos 251748 and 284621), Japan Society for the Promotion of Science (JSPS) Grant-in-Aid for Young Scientists (B) (No. 24740254), JSPS Postdoctoral Fellowships for Research Abroad, and JSPS Grants-in-Aid for Scientific Research (Grant No. 26400422). In addition, CSC--IT Center for Science Ltd. (Project No. ay2090) and Aalto Science-IT Project are acknowledged for computational resources.
\end{acknowledgments} 

%\section*{Acknowledgment}
% MN is grateful to JSPS for partial support from
%Grants-in-Aid for Scientific Research (Grant No. 26400422).

\end{document}